# Crystal Ball : On the Future High Energy Colliders

**Vladimir SHILTSEV**
*Fermilab* [1]
*PO Box 500, Batavia IL, 60510, USA*
*E-mail:* `shiltsev@fnal.gov`

High energy particle colliders have been in the forefront of particle physics for more than three decades. At present the near term US, European and international strategies of the particle physics community are centered on full exploitation of the physics potential of the Large Hadron Collider (LHC) through its high-luminosity upgrade (HL-LHC). A number of next generation collider facilities have been proposed and are currently under consideration for the medium- and far-future of the accelerator-based high energy physics. In this paper we offer a uniform approach to evaluation of various accelerators based on the feasibility of their energy reach, performance reach and cost range. We briefly review such post-LHC options as linear *e+e-* colliders in Japan (ILC) or at CERN (CLIC), muon collider, and circular lepton or hadron colliders in China (CepC/SppC) and Europe (FCC). We conclude with a look into ultimate energy reach accelerators based on plasmas and crystals, and some perspectives for the far future of accelerator-based particle physics.

*The European Physical Society Conference on High Energy Physics*
*22–29 July 2015*
*Vienna, Austria*

---

[1] Fermi Research Alliance, LLC operates Fermilab under contract no. DE-AC02-07CH11359 with the U.S. Department of Energy; supported, in part, by the European Commission under the FP7 EuCARD-2, grant agreement 312453

# 1.Introduction: Approach

Particle accelerators have been widely used for physics research since the early 20[th] century and have greatly progressed both scientifically and technologically since then. It is estimated that in the post-1938 era, accelerator science has influenced almost 1/3 of physicists and physics studies and on average contributed to physics Nobel Prize-winning research every 2.9 years [1]. Since the 1960's, twenty nine colliding beam facilities which produce high-energy collisions (interactions) between particles of approximately oppositely directed beams reached operational stage [2]. Their energy has ever been on average increasing by a factor of 10 every decade till about mid'1990's. Notably, the hadron colliders were 10-20 times more powerful. Since then, following the demands of high energy physics (HEP), the paths of the colliders diverged to reach record high energies in the particle reaction. The Large Hadron Collider (LHC) was built at CERN, while new *e+e-* colliders called "particle factories" were focused on detail exploration of phenomena at much lower energies. The Tevatron, LEP and HERA established the Standard Model (SM) of particle physics. Current landscape of the high energy physics is dominated by the LHC at CERN. The next generation of colliders is expected to explore it at deeper levels and to eventually lead the exploration of the smallest dimensions beyond the current SM.

Development of the energy frontier colliders over the past five decades initiated a wide range of innovation in accelerator physics and technology which resulted in 100-fold increase in energy (for each hadron and lepton colliding facilities) and $10^4$-$10^6$ fold increase of the luminosity. At the same time, it is obvious that the progress in the maximum c.m. energy has drastically slowed down since the early 1990's and the lepton colliders even went backwards in energy to study rare processes – see, e.g., Fig.1 in [3]. Moreover, the number of the colliding beam facilities in operation has dropped from 9 two decades ago to 5 now (2015). In this article we briefly review several future collider options which can be schematically bunched in three groups: *"near future"* facilities with possible construction start within a decade - such as international *e+e-* linear collider in Japan (ILC) [4] or circular *e+e-* colliders in China (CepC) [5] and Europe (FCC-ee) [6]; *"future"* colliders with construction start envisioned in 10-20 years from now – such as linear *e+e-* collider at CERN (CLIC) [7], muon collider [8], and circular hadron colliders in China (SppC) [5] and Europe (FCC-pp) [6]; and an ultimate *"far future"* collider with time horizon beyond the next two decades [3].

Discussion on the options for the future HEP accelerators usually comes to the question of right balance between the physics reach of the future facilities and their feasibility [3, 9, 10]. The concept of feasibility is quite complex and below we will attempt to offer a uniform approach to evaluation of various post-LHC colliders based on the feasibility of their energy reach (whether it is possible to reach the design c.o.m. energy of interest), feasibility of the performance reach (how challenging is declared design luminosity) and feasibility of the cost (is it affordable to build and operate?). While the first two criteria (energy and performance reach) are relatively easy to address on the base of the current state-of-the-art accelerator technology (of, e.g., superconducting magnets, RF, etc) and beam physics, the feasibility of the cost requires analysis of both the perspective available resources and the facility cost range.

Affordable cost of the frontier facility is crucial. As of today, the world's particle physics research budget can be estimated to be roughly 3B$ per year. Under the assumption that such financial situation will not change by much in the future and that not more than 1/3 of the total budget can be dedicated to construction of the next energy frontier collider over approximately a decade, one can estimate the cost of a globally affordable future facility to be about or less than 10B$ (in today's prices).

## 1.1 On the cost of frontier accelerators

An analysis of the known costs of large accelerator facilities has been undertanken in [11]. On base of publicly available costs for 17 large accelerators of the past, present and those currently in the planning stage it was shown that the "total project cost (TPC)" (sometimes cited as "the US accounting") of a collider can be broken up into three major parts corresponding to "civil construction", "accelerator components", "site power infrastructure" and the three corresponding cost components can be parameterized by just three parameters – the total length of the facility tunnels $L_f$, the center-of-mass or beam energy $E$, and the total required site power $P$ - and over almost 3 orders of magnitude of $L_f$, 4.5 orders of magnitude of $E$ and more than 2 orders of magnitude of $P$ the so-called "$\alpha\beta\gamma$-cost model" works with ~30% accuracy:

$$\text{Total Project Cost} \approx \alpha \times (Length/10km)^{1/2} + \beta \times (Energy/TeV)^{1/2} + \gamma \times (Power/100MW)^{1/2} , \quad (1)$$

where coefficients $\alpha=2B\$/(10\ km)^{1/2}$, $\gamma=2B\$/(100MW)^{1/2}$, and accelerator technology dependent coefficient $\beta$ is equal to *10 B$/TeV$^{1/2}$* for superconducting RF accelerators, *8 B$/ TeV$^{1/2}$* for normal-conducting ("warm") RF, *1B$/TeV$^{1/2}$* for normal-conducting magnets and *2B$/TeV$^{1/2}$* for SC magnets (all numbers in 2014 US dollars) – see Fig.1. Table 1 presents main parameters of the future colliders under discussion and their estimated costs.

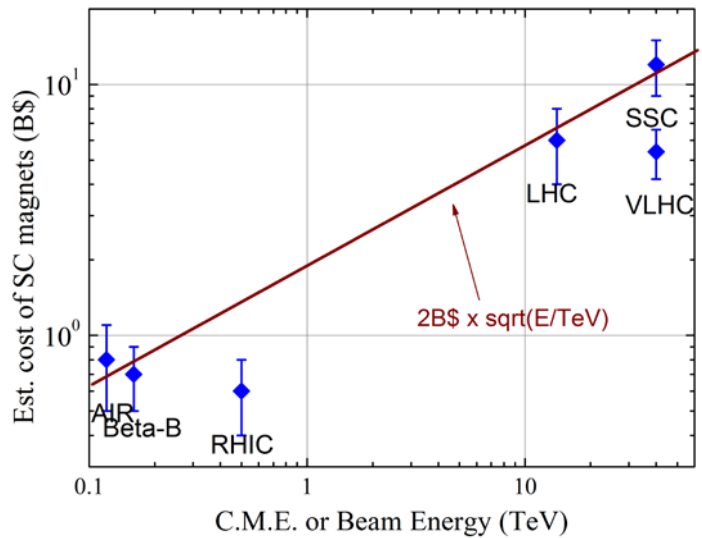

Fig. 1: Estimated cost of the SC magnets and associated elements vs collider center of mass energy or single beam energy. Stage I of the VLHC assumed low-field 2T superferric magnets (from [11]).

Very high total costs of the energy frontier colliders usually call for maximum possible performance (luminosity, see below), various measures to reduce the cost (extensive R&D on the cost-effective magnets and tunneling, re-use of the existing infrastructure and accelerator as injectors, etc.) and often the expansion of the physics program beyond primary colliding species (e.g., RHIC and LHC collide ions as well as protons). It also has to be noted that there are significant regional differences which should be taken into account. Sometimes they are indicative of the different methodology of the cost estimates – e.g., the "European accounting" includes only the industrial contracts for major items like civil engineering, the accelerator elements and corresponding labor requirements (such approach is often referred) and usually is factor of 2.0–2.5 lower than the US DOE Office of Science's "the total project cost" (TPC) which additionally includes the costs of the required R&D, development of the engineering design, project management, escalation, contingency, overhead funds, project-specific facility site development, sometimes — detectors, etc. Another notable difference is significantly lower cost of doing business in Asia, particularly, in China – for example, comparison of modern synchrotron light sources shows factor of about 3 lower construction cost for comparable facilities [12]. That advantage may or may not be in effect in the future but it definitely should be taken into account.

Table 1: Main parameters (c.m.energy $E_{cm}$, facility size $L_f$, site power $P$, luminosity $L$) of the collider projects under discussion and their estimated total project cost *TPC* according to the $\alpha\beta\gamma$–model [11] (see text).

|  | $E_{cm}$, TeV | $L_f$, km | $P$, MW | Region | $\alpha\beta\gamma$–TPC, $B (est.) | $L$, cm$^{-2}$s$^{-1}$ | Feasibility of Energy | Feasibility of Luminosity | Feasibility of Cost |
|---|---|---|---|---|---|---|---|---|---|
| *"Near" Future* | | | | | | | | | |
| CepC | 0.25 | 54 | ~500 | China | 10.2 ±3 | 5·10$^{34}$/IP | Y | Y? | Y |
| FCC-ee | 0.25 | 100 | ~300 | CERN | 10.9 ±3 | 5·10$^{34}$/IP | Y | Y? | Y? |
| ILC | 0.5 | 36 | 163 | Japan | 13.1 ±4 | 2·10$^{34}$ | Y | Y? | Y? |
| *Future* | | | | | | | | | |
| CLIC | 3 | 60 | 589 | CERN | 27.0 ±8 | 5·10$^{34}$ | Y? | Y? | N |
| μμ Collider | 6 | 20 | 230 | US ? | 14.4 ±5 | 2·10$^{34}$ | Y | N(yet) | Y? |
| SppC | ~50 | 54 | ~300 | China | 25.5 ±9 | 5·10$^{34}$ | N(yet) | N(yet) | N? |
| FCC-pp | 100 | 100 | ~400 | CERN | 30.3 ±9 | 5·10$^{34}$ | N(yet) | N(yet) | N? |
| *"Far" Future* | | | | | | | | | |
| X-Collider | ≤ 1000 | ≤ 10 | ≤ 100 | ? | ≤ 10 | 10$^{30-32}$ | ? | ? | ? |

## 2. Discussion: Frontier Accelerator HEP Facility Options

All three "near future" colliders are based on well developed technologies of NC magnets and SC RF and from that point of view their ability to reach the required c.o.m energies ("energy feasibility") has no seriuos doubts. The feasibility of performance with $L\sim(2\text{-}5)\cdot10^{34}$ cm$^{-2}$s$^{-1}$ per IP is not fully guaranteed becuase of a number of challenges, such as extraordinary overall facility power consumption (300-500 MW), heat load due to HOM heating in the cold SC RF cavities and beamstrahlung-limted dynamic aperture for circular *e+e-* colliders CEPC and FCC-ee [13], and the beam emittance generation and preservation in the main linacs and positron production for the ILC. All three facilities are on the brink of financial feasibility if the latter is defined at the TPC of 10B$ (note, that the publicly announced cost estimate of the ILC is 7.8B$ and 13,000 FTE-years of labor in the "European accounting" [4]).

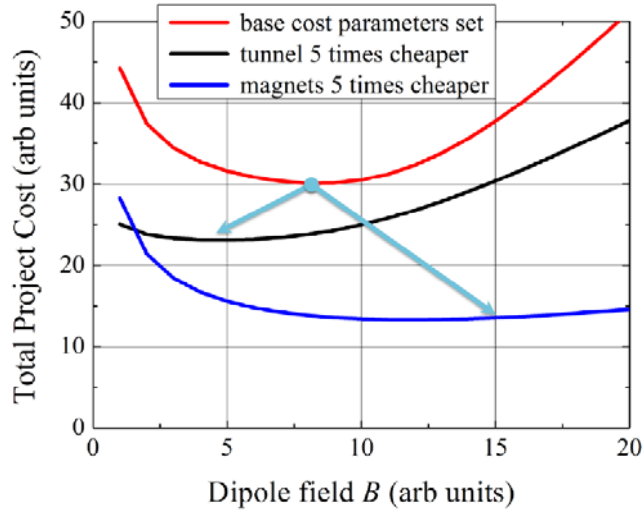

Fig.2: Total project cost of a Future Circular Collider vs maximum SC bending dipole field: red – the base cost parameter set per the $\alpha\beta\gamma$-model [11], black – in the case of the 5-fold reduction in the tunnel cost, blue – in the case of the 5-fold reduction of the SC magnet cost per Tesla-meter (for illustration only).

Among the ("medium") future colliders, only muon collider is based on the established technology of SC magnets and SR RF and, therefore, can fully guarantee the energy reach of upto 3-6 TeV c.o.m. It also seems relatively cost-effective and potentially affordable – see Table 1. Unfortunately, at the present state of the accelerator R&D program the performance of the muon collider can be assured at the level two to three orders of magnitude below the design luminosity goal of $2 \cdot 10^{34}$ cm$^{-2}$s$^{-1}$ and the performance feasibility requires convincing demonstration of the 6-D ionization cooling of muons [14]. The MICE experiment at RAL expected to provide the first experimental evidence of the muon cooling by 2018 [15]. Feasibility of the 3 TeV energy reach of the CLIC collider based on the novel two-beam acceleration scheme in 12 GHz normal conducting RF structures has only recently been demonstrated in a small scale CTF3 test facility where average accelerating gradients of 100 MV/m were achieved with acceptable RF cavity breakdown rates [7]. The luminosity goal of CLIC $L=5 \cdot 10^{34}$ cm$^{-2}$s$^{-1}$ is significantly more challenging than that of the ILC, though the design report indicates no principal showstoppers. The biggest issue for CLIC is its enormous site power consuption of about 600 MW and anticipated cost which probably can not be currently considered as affordable – see Table 1. Even a six-times smaller version of a 0.5 TeV c.o.m. *e+e*-collider based on the CLIC technology has been found quite expensive at 7.4-8.3BCHF and 14,100-15,700 FTE-years of labor [16]. Finally, the proton-proton supercolliders such as FCC-hh and SppC can not be claimed as "energy-feasible" as they require development of ~16T SC magnets which are at the edge of the reach of not-yet-fully developed $Nb_3Sn$ superconductor technology. Their required luminosity target of above $5 \cdot 10^{34}$ cm$^{-2}$s$^{-1}$ is not achievable until critical issues of the synchrotron radiation heat load in the cold magnets, machine protection, ground motion and many others are addressed [12, 17, 18]. The biggest challenge of such huge machines with 60 to 100 km circumferences is their cost. Indeed, according to the $\alpha\beta\gamma$-model Eq.(1), the cost of 100 km long accelerator facility with some 400MW of site power and based on today's SC magnets can be estimated as *TPC=2×(100/10)$^{1/2}$+2×(100 TeV/1TeV)$^{1/2}$+2×(400/100)$^{1/2}$ =30.3B$±9B$.* As the biggest share of the TPC is for the magnets, the primary goal of the long-term R&D program should be development of ~16T SC dipole magnets which will be significantly (by a factor 3-5) more cost effective per TeV (or Tesla-meter) then those of, say, LHC – see Fig.2.

While talking about frontier colliders, one should take into account the availability of experts. A simple "rule of thumb" (also know as "Oide-principle" [19]) based on statistics of construction projects in Japan and Europe and widely accepted in the accelerator community states that "one accelerator expert can spend intelligently 1 M$ in one year". One can estimate that the world-wide community of accelerator physicists and experienced engineers does not exceed 1500 people and the total accelerator personnel (all scientists, engineers, technicians, drafters, etc) is about 4,000-4,500. Therefore, any plans for a really big facility at the scale of few B$ to 10B$ should take into account that significant time will be needed to get the required number of the people together. Another comment deals with the fact that due to extremely cpmplex nature of the fronrtier accelerators it takes time to get to design luminosity - often as long as 3-7 years [20] – and that should also be taken into account in any realistic plans.

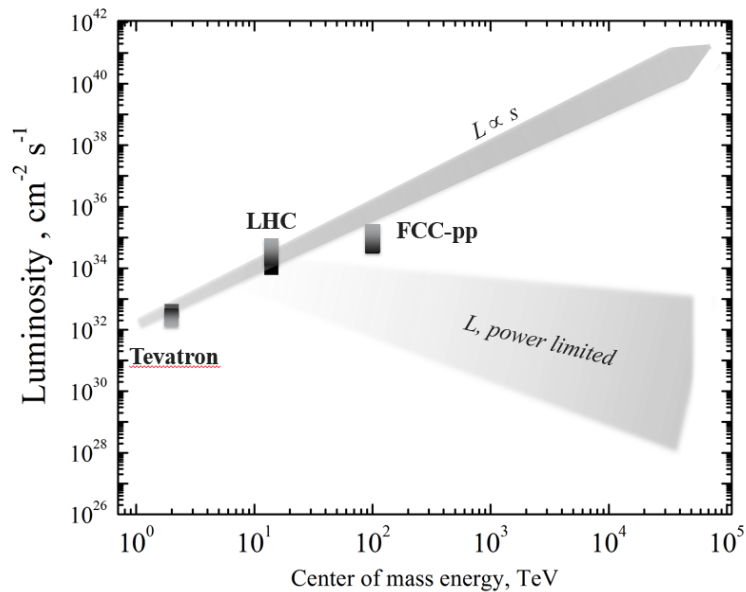

Fig.3: "Luminosity vs Energy" paradigm shift (see text)

Finally, one can try to assess options for "far future" post-FCC energy frontier collider facility with c.o.m. energies (20-100 times the LHC (300-1000 TeV). We surely know that for the same reason the circular *e+e-* collider energies do not extend beyond the Higgs factory range (~0.25 TeV), there will be no circular proton-proton colliders beyond 100 TeV because of unacceptable synchrotron radiation power – they will have to be linear. It is also appreciated that even in the linear accelerators electrons and positrons become impractical above about 3 TeV due to beam-strahlung (radiation due to interaction at the IPs) and about 10 TeV due to radiation in the focusing channel (<10 TeV). That leaves only $\mu+\mu-$ or *pp* for the "far future" colliders. If we further limit ourselves to affordable options and request such a flagship machine not to exceed ~10 km in length then we seek a new accelerator technology providing average gradient of >30 GeV/m (compare with $E/L_f$~ 0.5 GeV per meter in the LHC). There is only one such option known now: dense plasma as in, e.g., crystals, that excludes protons (due to nuclear interactions) and leaves us with muons as the particles of choice [3]. High luminosity can not be expected that such a facility is we limit the beam power and, with necessity, the total facility site to some affordable level of ~100MW. Indeed, as the energy of the particles $E$ grows, the beam current will need to go down at fixed power $I=P/E$, and cosequently, the luminosity will need to go down with energy – see Fig.3. The paradign shift from the past collider experience where $L \sim E^2$ will need to happen in the "far future" of HEP.

**3. Summary**

Summarizing the collider's past and present situation we can remark great success of the colliding beams method: 29 colliders were built over 50 yrs and energies of O(10) TeV c.o.m. achieved (LHC). The progress has greatly slowed down in the past two decades due to increasing size, complexity and cost of the facilities. Accelerator technologies of RF and magnets are well developed and costs understood (and can be paramterized by the $αβγ$-model).

Under thourough consideration are the "near future" facilities with construction start within next decade such as CepC, FCC-ee and ILC. They are are compelx but feasible in terms of energy, luminosity and, possibly, cost. CepC in might have "unfair competitive advantage" due to significantly lower construction costs in China. It is understood that such facilities will need ~700-1000 accelerator experts and getting such a team together will require significant time. One should also not expect the design luminosity of such colliders to be available on "day 1" - instead, more like in the "year 4 or 5" of operation.

Future energy frontier colliders with possbile construction start ~ 2 deacdes from present have serious issues: 3 TeV CLIC - with performance and cost, 6 TeV muon collider - with performance (luminosity), 60-100 TeV FCC-hh and SppC - with cost and performance. The key R&D area for the FCC-hh and SppC is developement of cost effective ~16T SC magnets. That might take take as long as two decades, but seemingly all three regions – the US, Europe and Asia - are open for start such collaborative R&S now.

Our brief outlook into the "far" future colliders shows that there are not so many options for the facility with energy reach 30-100 time the LHC. Actually, there is seemingly only one - linear acceleration of muons in dense plasma. In any case, such ultra-high energy collider be necessity will have low luminosity.

This talk was presented and discussed at the EPS conference on High Energy Physics held in Vienna, Austria, on July 22-29, 2015. I would like thank the conference organizers and, particularly, Jochen Schieck for the hospitality and the accelerator session conveners John Jowett and Frank Zimmermann for helpful practical assistance. I greatly appreciate support provided by the EuCARD-2 WP to attend the conference. Stimulating discussions which I had with Mei Bai, Alain Blondel, Fredrick Bordry, John Jowett, Ernie Malamud and Frank Zimmermann made this conference indeed remarkable.